\documentclass{sig-alternate-2013}

\usepackage[colorlinks]{hyperref}
\usepackage{breakurl}
\usepackage{listings}
\usepackage{graphicx}
\usepackage{amstext}
\usepackage{bbm}
\usepackage{amsfonts}
\usepackage{amssymb}
\usepackage{xspace}
\usepackage{multirow}

\renewcommand{\url}{\path}

\usepackage{color}

\newcommand{\eg}{e.g.\xspace}

\newcommand{\etc}{etc.\xspace}

\newcommand{\apriori}{\emph{a priori}\xspace}

\newcommand{\e}[1]{\ensuremath{\times 10^{#1}}}
\newcommand{\vect}[1]{\ensuremath{\mathbf{#1}}}
\newcommand{\abs}[1]{\ensuremath{\left\lvert#1\right\rvert}}
\newcommand{\norm}[1]{\ensuremath{\lVert#1\rVert}}

\DeclareMathOperator*{\argmin}{arg\,min}

\expandafter\def\expandafter\UrlBreaks\expandafter{\UrlBreaks \do\*\do\-\do\~\do\'\do\"\do\-}

\permission{} 
\copyrightetc{}

\begin{document}

\title{Malicious Behavior Detection using Windows Audit Logs}

\numberofauthors{3}

\author{
\alignauthor{Konstantin Berlin}\titlenote{Corresponding author.}\\
       \affaddr{Invincea Labs, LLC}\\
       \affaddr{Arlington, VA, USA}\\
       \email{kberlin@invincea.com}
\alignauthor{David Slater}\\
       \affaddr{Invincea Labs, LLC}\\
       \affaddr{Arlington, VA, USA}\\
       \email{david.slater@invincea.com}
 \alignauthor{Joshua Saxe}\\
       \affaddr{Invincea Labs, LLC}\\
       \affaddr{Arlington, VA, USA}\\
       \email{josh.saxe@invincea.com}
}

\maketitle

\begin{abstract}
As antivirus and network intrusion detection systems have increasingly proven insufficient to detect advanced threats, large security operations centers have moved to deploy end\-point-based sensors that provide deeper visibility into low-level events across their enterprises. Unfortunately, for many organizations in government and industry, the installation, maintenance, and resource requirements of these newer solutions pose barriers to adoption and are perceived as risks to organizations' missions.

To mitigate this problem we investigated the utility of agentless detection of malicious endpoint behavior, using only the standard build-in Windows audit logging facility as our signal. We found that Windows audit logs, while emitting manageable sized data streams on the endpoints, provide enough information to allow robust detection of malicious behavior. Audit logs provide an effective, low-cost alternative to deploying additional expensive agent-based breach detection systems in many government and industrial settings, and can be used to detect, in our tests, $83$\% percent of malware samples with a $0.1$\% false positive rate. They can also supplement already existing host signature-based antivirus solutions, like Kaspersky, Symantec, and McAfee, detecting, in our testing environment, $78$\% of malware missed by those antivirus systems.
\end{abstract}

\section{Introduction}

Recent, public, high-profile breaches of large enterprise networks have painfully demonstrated that cybersecurity is rapidly becoming one of the more daunting  challenges for large organizations and government entities. The challenge stems from the relative ease with which malware authors can permute, transform, and obfuscate their cyberweapons in order to avoid signature based detection, the dominant approach for detecting cyber threats. 

One  strategy for combating obfuscation of the malware signature is to perform detection on the actual dynamic behavior of software, since obfuscating behavior is potentially harder, and requires significantly more research and development to create new behavior infection vectors \cite{moser2007limits}. Currently, collecting such dynamic behavior on an enterprise network is a huge challenge, primarily because it requires instrumentation of individual machines, or redirection of files to be run on a centralized virtual environment (\eg, FireEye, Lastline), which can be costly, require installation of third party software, and can be difficult to maintain. Finally, most newer systems are typically perimeter defenses, where the malware is analyzed in a virtual environment, and can do nothing in the case when software manages to by bypass it (\eg, by requiring a reasonably complicated user interaction in order to unpack itself and execute).

At the same time most organizations have existing infrastructure for monitoring their endpoint machines directly, called Security Information and Event Management (SIEM) systems, which collect vast amounts of security information from actual software being executed on an endpoint. This is an important distinction, since during actual execution, malware has no choice but to expose its behavior. Unfortunately, this information is typically used for forensic analysis, rather than detection. In the cases when detection is done, it is usually in the form of a network intrusion detection system (IDS), where the focus is to discover anomalous communication on the enterprise network, rather than malicious software behavior on a host \cite{yen2013beehive}. 

One of the low hanging fruits not currently utilized in dynamical behavior detection are the Windows audit logs. Windows audit logs can be easily collected using existing Windows enterprise management tools \cite{audit2015}, and are built into the operating system, so there is a low performance and maintenance overhead in using them. In our work we investigate the potential of these audit logs to augment existing enterprise defense tools. Our main contributions are as follows:
\begin{itemize}
\item We demonstrate a scalable computational approach that processes audit logs into an interpretable set of features, which can then be used to build a malware detector.
\item We demonstrate that a linear classification model, using a small subset of audit log features, is able to detect malicious behavior in a simulated enterprise setting with high accuracy.
\item We show that this classifier can provide significant detection of malware that are completely missed by top antivirus vendors. Thus providing a cheap value-add to already installed host-based systems.
\item We describe some interesting malware behavior that we discovered to be significant signals of malware, which supports our claim that our detections works on malicious behavior rather than behavioral signatures.
\end{itemize}

The rest of this manuscript is organized into several major sections. We start with Section \ref{sec:background}, where we present relevant background information relating to malware detection techniques, and machine learning. Then, in Section \ref{sec:experimental}, we describe the dataset of audit logs that we use to compute and evaluate our machine-learning classifier on, as well as how we determined classification labels for these logs. In Section \ref{sec:method} we describe how we extracted relevant features from the collected logs, and based on these extracted features, how we learn our classifier. We show our results in Section \ref{sec:result}, and end with a conclusion in Section \ref{sec:conclusion}. 

\section{Background}
\label{sec:background}

Malware detection can be roughly split into two major approaches: i) detection based on static analysis, where the suspected executable file is scanned for malicious content; and ii) dynamic behavior detection, where the detection is done on the behavior of the executable as it is running \cite{egele2012survey}.

In the commercial antivirus industry, static analysis has been the dominant approach for detection, and is typically implemented by blacklisting malware executables based on their byte signatures.  Blacklisting using signatures is very effective in actual deployment, since, when properly done, the signature detector can have virtually zero false positives, with detection rates $\geq 90$\% for common malware\cite{mohaisen2014av,avcompare2015}.

Unfortunately, since detection is done on signatures, small modifications of the source code, compiler, or the binary would evade the signature method. Even more advanced detection methods are still susceptible to simple obfuscation techniques \cite{moser2007limits}. Since small modifications of malware can be done at a low cost, the result has been an explosion in new malware being observed in the wild, and a significant lag between when a piece of malware is first observed and the time antivirus engines are able to detect it \cite{lastline2015}.

A complementary approach to static detection is dynamic behavior detection. Behavior is more difficult to obfuscate, so the advantage of dynamic analysis is that it is harder to recycle existing malware. Software is typically observed by automatic execution in virtualized environments (\eg, \cite{cuckoo2015}), prior to allowing it to run on the endpoint system, in order to prevent infection. Automated execution of software is difficult since malware can try to detect if it's running in a sandbox, and prevent itself from performing malicious behavior, which essentially leads to an arms race between dynamic behavior detectors using a sandbox and malware \cite{anubis2015,fleck2013pytrigger}. A somewhat related concept to dynamic tracing is taint analysis \cite{yin2007panorama,enck2014taintdroid}, which requires significant instrumentation to the underlying system and can suffer from its own set of issues \cite{slowinska2009pointless}.

While we are proposing a dynamic behavior detection system, our approach is complementary to the above described methods, since the detection is done after the malware bypasses perimeter defense layers (like antivirus and sandbox) and executes behavior directly on the endpoint, where it has no choice but to behave maliciously if it is to accomplish its goal. Instead of proposing a custom agent-based approach, our work is motivated by the practical issues that typically prevent deployment of any new tools, no matter how effective they might be, such as reticence of IT departments to install new software, additional maintenance requirements, and costs. In our approach we mitigate those issues by taking advantage of already existing Windows audit logs collection mechanisms in most SIEMs. While such an approach limits the type of event data that we can use to what is currently recorded by Windows audit logs, collection of his data can be implemented on a network administration level without any software installation, requires no additional maintenance of the hosts, and, as we describe later, induces minimal network and system overhead.

\subsection{Machine Learning}

Machine learning has been used extensively in static and dynamic malware detection (\eg, see a recent review \cite{gandotra2014malware}), though the potential performance in an enterprise setting is hard to judge, since results are typically computed on different, small datasets. The detection problem is commonly formulated as either a classification or an anomaly detection problem, depending on the type of data that is used. In anomaly detection, malicious behavior is characterized deviation from the ``normal'' expected behavior. The advantages of such an approach is that only normal behavior needs to be collected for training, which is very easy to get in large volumes on an enterprise network. In cybersecurity, anomaly detection methods are typically applied to network intrusion detection problems, where network flow traffic is analyzed \cite{yen2013beehive}.  

Trying to detect abnormal behavior can be problematic when dealing with active users, since behavior can periodically deviate from expectation. For example, when new software is installed, job assignment changes, or a new user takes over a computer temporarily, the audit logs can deviate from expectation. In an enterprise setting, false alarms can have a very detrimental effect, since it can cause the administrators to lose faith in the detector. In this situation, we can potentially improve detection by also utilizing malware-associated signals. We therefore focus on expressing audit log malware detection as a classification problem, in the hopes that this will yield a more robust solution. 

\section{Experimental Data}
\label{sec:experimental}

In this section we describe how we setup the Windows Audit Log to collect the required data, as well as the actual dataset that we will use for learning and evaluation. We start by describing our Audit Log settings and the types of behavior that we record. We then describe our efforts to collect real enterprise audit logs using these settings, as well as a set of diverse software samples (benign and malicious) that we run through a virtual sandbox environment. Finally, we discuss how we decide which binaries are benign and which are malicious for use in the training and testing of our detection approach.

\subsection{Configuration}

Collecting Windows audit logs, while straightforward, requires defining the type of system objects that will be monitored (\eg, files or registry keys, network events), what types of access to those objects should be recorded (reads, writes, permission changes, \etc), and whose accesses should be monitored (users, system, all, \etc). The configuration for this information is held in three places: i) the security access control list (SACL), which controls the access types audited for all file and registry objects; ii) the local security policy, which controls what sorts of objects and events generate events in the log; and iii) the Windows firewall, which partially controls logging for network specific events, such as network connections. The specific policy can easily be applied to the entire enterprise network using domain controllers.

While turning on all logging mechanisms would provide the most coverage of possible behaviors, the volume of log events quickly overwhelms commodity storage systems and significantly impacts the performance of the monitored computers. To filter out the vast majority of these events while maintaining their utility in detecting malicious behavior, we removed read events from our collection, as nearly all malicious behavior involves some form of modification to the system (reconnaissance is one notable exception, though the software still has to gain permanence). We also did not record network events, due to the limitation of our sandbox environment. In total, we end up only recording file/registry writes, deletes, and executes, and process spawns. Our settings generated about 100-200 MB of data (representing about 300-600 thousand events) per computer per day on an enterprise network, with  peak volume days being about double that. The logs can be compressed down for long-term storage at a rate of 16:1.  No detectable performance degradation was detected by any users of the monitored machines. Thus, activating the additional logging needed for our approach does not pose a heavy storage or network load for a modern enterprise network, and has negligible memory or computational overhead.

\subsection{Collection}

Our collected audit logs consists primarily of two sources: i) the enterprise logs collected from users of an enterprise network; and ii) sandboxed virtual machine based runs on a set of malicious and benign binaries. While the enterprise logs represent exactly the environment on which we propose to deploy our system, we are unable to collect any malicious behavior that we require for machine learning. 

The enterprise logs we collected came directly from an internal enterprise network, recorded using the described audit log configurations. These logs consist of four Windows computers actively used by members of our sales, executive, and IT teams.

To generate malicious audit logs, as well as to diversify the benign behaviors recorded, we created a virtual sandbox environment in which we run several collections of binaries. We used Cuckoo Sandbox (CuckooBox) \cite{cuckoo2015}, an open source sandbox, in conjunction with VirtualBox \cite{virtualbox2015} running Windows 7 SP1 VMs to automatically generate these audit logs from the binaries. We ran about twenty thousand binary samples through CuckooBox, with each sample taking an average of four minutes per run, with the maximum allowable runtime set to ten minutes. We will collectively refer to this dataset of logs generated by running through CuckooBox as the CuckooBox dataset. The binaries were sampled from our collection of portable executable (PE) format Windows files:
\begin{itemize}
\item (MAL2M) Over two million malware samples collected in 2012. While majority of the files are malware, some small fraction is incorrectly labeled benignware.
\item (MAL3P) Around one thousand malware binaries manually created by a third party. All these files are known to be malware.
\item (MALAPT) Five sets of binaries recovered from high profile APT cyber incidents. All these files are known to be malware.
\item (UVPN): Around sixteen thousand binaries download by virtual private network users that we received from our corporate partner. These binaries consist of downloaded files sampled from several million active end users, providing a recent and realistic set of mixed malicious/benign binaries, allowing us to estimate real world binary execution behavior. The set contains a mix of benign and malware executables. 
\item (OS) Windows XP Service Pack 3 and ReactOS (open source Windows clone) system files. All these files are known to be benignware.
\end{itemize}

Note that our CuckooBox dataset of audit logs contains both malware and benign audit logs.

\subsection{Labeling}

\begin{figure}[ht]
	\center
	\includegraphics[width=0.5\textwidth]{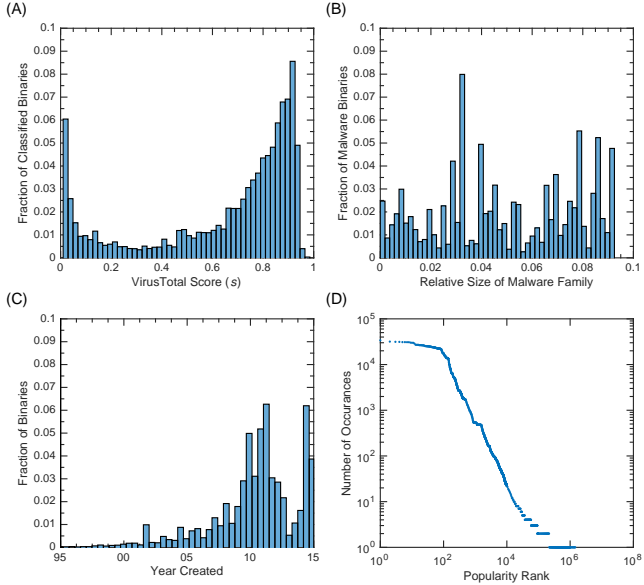}
   \caption{Characterization of our database binary files and logs. (A) The VirusTotal classification of all the binaries ran through CuckooBox with $s>0$. (B) The distribution of our malware binaries as a function of family size.  (C) The distribution of compile timestamp (or file creation data, if timestamp is unavailable) in our CuckooBox run binary files. Extreme values (due to corrupted stamps) are not shown. (D) The popularity of events produced in CuckooBox runs of binaries, ranked by the number of occurrences (where events are defined by action type and target). This demonstrates a long tail of observed malicious events. }
   \label{fig:data}
\end{figure}

In order to use classification algorithms to build a detector we must provide classification labels of ``malware'', $1$, or ``benign'', $-1$,  for all of our samples. Given the non-negligible chance of benignware occurring in the malware MAL2M dataset, and lack of any \apriori classification of the UVPN binaries, we ran all the used binaries through VirusTotal. VirusTotal uses approximately $55$ different malware engines to see if any of them label the executable as malware \cite{virustotal2015}. We will define the VirusTotal score $s$, where $0\leq s \leq 1.0$, as the number of detections divided by the number of malware engines. The distribution of scores on our binary dataset is shown in Fig. \ref{fig:data}B.

To reduce the chance of mislabeling we remove all binaries with ambiguous scores of $0 < s < 0.3$, based on the valley centered at around $s=0.3$ in Fig. \ref{fig:data}A. Any binary with a score of $s=0$ we label as benignware, and $s\geq0.3$ we label as malware. The exception are any binaries for which we know the correct labels based on the data source: MAL3P and MALAPT we always label as malicious; and OS binaries we always label as benign.

Our CuckooBox audit log dataset contains 981 different malware families, as characterized by Kaspersky antivirus. As shown in Fig. \ref{fig:data}B, our corpus is well distributed over the size of the malware family, it's VirusTotal score (indicating whether it is considered malicious or benign), its creation time (with some very recent ones coming from VPN dataset), and the relative popularity of specific audit log events produced when running a binary. In particular, relative popularity is ranked by the number of times a given event occurred, where an event is represented by a string tuple $\langle a,b\rangle$, where $a$ is the action (\eg, spawn, write, \etc) and $b$ is the associated file path or regularized path. Note that this is not the popularity of the malware binary itself. 

Out of $14679$ confirmed malicious binaries that we successfully ran through CuckooBox, Kaspersky detected $88$\% of malware, McAfee $95$\%, and Symantec $90$\%. If we combine the above engines into a ``meta''-engine, where we consider a detection valid if at least one of the engines detects malware, then for McAfee+Symantec we have $96$\% detection, and for McAfee+Symantec+Kaspersky we have $98$\% detection.

Note that our detection percent for various antivirus engines is potentially more optimistic than in real deployment environment for several reasons: (i) a large fraction of our malware is more than two years old; ii) while we potentially have newer malware in the VPN binary dataset, there is a chance that it is not detected by more than  $30$\% of antivirus engines, and is so is not included in our statistics; and (iii) the antivirus engines are potentially setup with more aggressive detection settings than they would otherwise be when actually deployed.

In the rest of the manuscript we will refer to $\vect{y}$ as the classification vector of size $M$, where $y_i\in \{-1,1\} $ is the $i$th sample's classification,  $M$ is the number of observations in the dataset, and $-1,1$ means benign or malicious, respectively.

\section{Method}
\label{sec:method}

Our method for deriving a classifier from our audit logs is divided into two stages. In the first stage we perform the feature engineering, where we use domain knowledge to extract features out of audit logs. In the second stage we use machine learning to compute a classifier that classifies audit logs based on the extracted features.

\subsection{Feature Engineering}

Our malware logs, on average, are $4$ minute long recordings of the whole system during which a binary was executing in CuckooBox. On the other hand, the enterprise logs are continuous recording lasting for hours at a time. We therefore split all our enterprise logs into disjoint $4$ minute windows to mimic our CuckooBox runs. In order to remove some common host specific artifacts from file paths and registries, we ran them all through a series of regular expression that abstracted out the host specific strings (ex. username in the path, UIDs in the registry, \etc).

The most direct approach for mapping events in the log to features is to represent each single event as a single feature, and represent each log as a bag-of-words of individual events. However, the time order between events is lost in such a mapping. This potentially might not significantly impact results, since malware can be multithreaded, and infect multiple processes at the same time, so it is not possible to unambiguously order all log events. However, some orderings have meaning, since an event could represent different behavior depending on the context.

One improvement over the one-to-one mapping, which partially preserves order, is to represent all $q$ continuous time ordered sets of events (event $q$-grams) in the log as features, for some number $q$. For example, one event could be execution of ``a.dll'', followed immediately by execution of ``b.dll'', rather than two independent features. The drawback of such an approach is that it causes an exponential increase in the number of features, limiting how big we can computationally make $q$. Also, as we make $q$ larger we are potentially getting features that do not generalize as well, so the utility of computing $q$-grams for large $q$ is low. In our approach, we group all events in a log by their associated process IDs, extract all $q$-grams for each process, and then aggregate them together into one large feature set of $q$-grams that represents the entire log. We set $q=\{1,2,3\}$, a range of sizes we determined to be a practical compromise between computational complexity and goodness of results. Representing all events of 1-3 $q$-grams resulted in about seven million unique features. 

We can now represent our combined benign and malware audit log dataset simply by an $M\times N$ matrix $\vect{A}$, where entry $a_{ij}$ is the number of observations of feature $j$ in log $i$. Unfortunately there is a danger that different users have a different number of benign processes running at a time, so the counts can vary significantly between different logs within the same timeframe. In addition, the amount of events occurring in CuckooBox is potentially smaller than on an enterprise user machine, where more software is actively being used. To reduce the effect of heterogeneity in our audit log data, we drop all counts from our matrix, such that the matrix only stores binary values, 0 if the feature never occurred in the log, and $1$ otherwise.

When running binaries in the sandbox it is possible that a fraction of these binaries did not execute properly due to various reasons. To filter them out, we compute the mean and standard deviation of the number of features for our CuckooBox audit log dataset, and removed all the observations where the number of features was below two standard deviations from the mean. 

\subsubsection{Feature Filtering}

The large number of features that we extract from the logs can make it intractable to compute a classifier. In our case, most features are not useful for detection because they occur in a tiny fraction of logs (see Fig. \ref{fig:data}D). Our first step is to filter out the long tail of the feature distribution, such that we would be able to tractably apply a standard machine learning algorithm. Our goal is to do a computationally fast filter that will preserve the sparsity of the data matrix. Therefore, we avoid performing expensive operations like matrix decompositions, and filter the features based on the uncentered correlation coefficient, $c_j$, between the labels $\vect{y }$ and the $j$th column of $\vect{A}$, $\vect{a}_{*j}$,
\begin{equation}
	c_j = \ \frac{\vect{a}_{*j}^T \vect{y}}{\norm{\vect{a}_{*j}}_2 \norm{\vect{y]}}_2},
\end{equation}
which can be efficiently computed using a sparse matrix multiplication. Using $\abs{c_j}$ values, we select the top $50000$ most correlated (or inverse correlated) features, a conservative number that is small enough to practically train our machine-learning classifier on a desktop computer, but two orders of magnitude bigger than the number of popular events in our log, while still preserving sparsity of the feature matrix.

Note that in the case when $N$ is extremely large (ex. billions), we can pre-filter our correlation filter by using a probabilistic counter \cite{mitzenmacher2005probability}, and performing two passes through the audit log dataset. On the first pass we count the number of occurrences, and on the second pass we only add a feature to $\vect{A}$ if is above a certain threshold. Fortunately, for our current dataset, we were able to compute the correlation directly without pre-filtering.

\subsection{Learning Algorithm}

Picking the optimal machine learning approach for a specific problem is a research topic itself, and so we leave building the most optimal detector to future work. In our case, we focus on building an easily deployable detector that can perform detection very quickly using very few important features, since: i) we can practically deploy such a detector in a large enterprise network by using simple pre-filtering for important features on the SIEM system, activating the more expensive detection system only in the case when one of the important features occurs; and ii) in the case when a detection does occur, we can provide an explanation that can be verified by a human agent. Therefore, we will focus specifically on the two class $\ell_1$-regularized logistic regression (LR) machine-learning classifier, where the $\ell_1$-norm induces a sparse solution \cite{tibshirani1996regression}. 

To confirm our results, which we give later, we have also tried SVMs and random forests \cite{shalev2014understanding}, with both methods yielding similar performance to LR. We do not report on those results.

\subsubsection{Logistic Regression}

An $\ell_1$-regularized LR classifier can be computed from $\vect{y}$ and $\vect{A}$ by minimizing the associated convex loss function
\begin{equation}
	\langle\hat{\vect{x}},\hat{b}\rangle = \argmin_{\vect{x},b}  \sum_i \log\left(1+e^{-y_i \left(\vect{a}_i\vect{x}+b\right)}   \right)  + \lambda \norm{\vect{x}}_1,
	\label{eq:lr}
\end{equation}
where $\vect{a}_i$ is the feature row vector of the $i$th observation, $y_i$ is the associated label, $\vect{x}$ is the column vector of the classifier feature weights, $b$ is the offset value, $\norm{\cdot}_1$ is the $\ell_1$-norm, and $\lambda$ is a regularization parameter. We will describe how we pick $\lambda$ through cross-validation in the next subsection.

Given the LR classifier $\langle\hat{\vect{x}},\hat{b}\rangle$, the probability of the $i$th observation being malware, $p_i$,  is evaluated using the logistic function
\begin{equation}
	p_i=\frac{1}{1+e^{-\left(\vect{a}_i\hat{\vect{x}}+\hat{b}\right)}}.
	\label{eq:p}
\end{equation}
 
\subsubsection{Regularization}

The value of the regularization parameter $\lambda$ is not known \apriori, since it unclear what the expected total loss is for our dataset. In order to determine the proper $\lambda$ parameter in eq. \eqref{eq:lr} we will use internal $10$-fold cross-validation, where for each $\lambda$ value we compute the average loss of the validation sets. The cross-validation can be computed fairly quickly, since the optimization problem can be sped up using warm restarts, where we use the solution from a previously computed $\lambda$ value to converge faster to the solution for the new $\lambda$ \cite{friedman2regularization}. Rather than picking the value of $\lambda$ that gives the lowest loss during cross-validation, we will err on the side of parsimony and use the ``one-standard-error'' rule, since our deployment environment will somewhat diverge from our testing environment \cite{friedman2regularization}. Note that this validation is not our actual validation, which we report on later, but an internal cross-validation to select the $\lambda$ value.

\subsection{Validation}

The LR classifier's tradeoff between detection and false alarm rates can be controlled by the threshold value at which we consider the probability $p_i$ (see eq. \eqref{eq:p}) high enough to be malicious. The full range of this tradeoff is summarized by the receiver operating characteristic (ROC) curve, where on the $y$-axis we have the true positive rate (TPR), the number of malicious logs classified by our classifier as malicious, divided by the total number of malicious logs; and on the $x$-axis we show the false positive rate (FPR), the number of logs we classified as malicious that are actually benign divided by the total number of benign logs. We evaluate our approach based on TPR and FPR because both are independent of the ratio of benign vs. malicious logs in our dataset, which varies between deployment environments.

The TPR and FPR cannot be properly estimated using the data that was used to learn the classifier, and instead must be computed through validation on previously unseen data, where the classifier is trained on ``training'' data and tested on ``test'' data, with the most common type of validation being cross-validation \cite{shalev2014understanding}. Unfortunately, direct application of the random split cross-validation approach, in our case, could provide misleading results due to the differences between our validation data and the deployment environment. This is a direct consequence of us not being able to observe malware infections directly on an enterprise network, and so having to simulate the infections using a sandbox. 

Consider a typical enterprise deployment environment: on a typical day there are a multitude of users active, most of them using Office products, a web browser, or conferencing applications; lots of software is actively running on each computer, and is being constantly interacted with, but software installations are uncommon. We contrast this with the CuckooBox environment that we used to exhibit malware behavior, where only a few processes are active, there is no active interaction with any applications, but software is constantly being installed and CuckooBox monitoring tools, which would not exist on a real enterprise machine, are mixed in with actual monitored software behavior. Such heterogeneity between benign enterprise data and malware CuckooBox data can result in us not being able to distinguish between a classifier that primarily detects CuckooBox vs. enterprise environment and one that actually detects malicious behavior.

Another problem in estimating the TPR and the FPR from validation results is the unexpected dataset ``twinning'',  where a fraction of our samples in the test set are behaviorally almost identical to the samples in the training set. Indeed, we have observed executions of binaries with different SHA1s that produced almost identical audit logs. Related to this, is the problem of concept drift, where malware behaviors tend to evolve over time and drift away from the samples in our dataset. In both cases, random split validation would produce overly optimistic results.

On the other hand, if our classifier is actually classifying based on behavior, estimating the FPR only on CuckooBox benignware might result in a pessimistic rate, since we potentially have a number of malware binaries in our VPN dataset that were missed by VirusTotal meta-engine, but are detected by our classifier. Additionally, separating behavior of VPN binaries (which mostly contain new software installers) from malware is much harder than separating typical behavior of an enterprise endpoint from malicious behavior, since those tend to deviate less from expected behavior. 

Therefore, in order to present a more accurate estimate of expected performance, we have designed six different validations that, when taken together,  show that our classifier provides robust performance under the more realistic set of assumptions. First, to address dataset twinning or concept drift, in addition to the validation based on random dataset splitting, we also validate results by testing only on malware that is at least one or two years older than malware in our training set. Since compile time in an executable can be faked or be corrupted, we remove all executables that have compile time before the year 1995 or compile time after 2014.

As an alternative, we also validate by splitting on malware families instead of compile time, where we define a family based on Kaspersky's label. In our malware families test we remove ``Trojan.Win32.Generic'' from both the training and testing datasets, though we acknowledge that family labeling can be somewhat ambiguous \cite{mohaisen2014av}. 

The second variable in our validations is the environment of our test data, where we test the same classifier, first in CuckooBox, and then enterprise environments, separately. By maintaining the environment constant for the benign and malicious samples in the test set, we (mostly) mitigate the possibility that our evaluation results are tainted due to the heavily biased ratio of benign vs. malicious samples in our two environments. We describe this approach and how we generate malicious samples for the enterprise environment next. 

\subsubsection{Mitigating Environmental Bias}

We start with the the closest approximation to the standard 10-fold cross-validation approach, where we split all the CuckooBox datasets into $10$ disjoint sets, and for each validation iteration train on the $9$ out of $10$ CuckooBox sets and all of the enterprise data, and then validate on the remaining CuckooBox set (no enterprise data). Note the difference between this approach and standard cross-validation, where we would also validate on the enterprise data. While this would demonstrate that our classifier does indeed distinguish between benignware and malware in the same environment, it is not clear what our performance is on actual enterprise data. If our FPR on the enterprise data is lower than the fraction of benignware that is mislabeled as malware, our reported FPR might be artificially high. 

One approach would be to get rid of the CuckooBox benign data, and instead test directly on CuckooBox malicious vs. enterprise datasets. However, testing directly on enterprise data and synthetically generated malware CuckooBox logs could be misleading, since it is still possible to separate the benign samples from malware samples just by performing detection for CuckooBox environmental features (\eg, CuckooBox specific DLL calls). It is also conceivable for a bad classifier to pass the CuckooBox validation, and also the proposed validation, by first detecting the execution environment, and if it does not detect the CuckooBox environment then automatically classify it as benign, otherwise perform actual detection using some CuckooBox model. We avoid such a possibility by synthetically generating enterprise {\it malicious} logs by logically OR-ing it with a malware CuckooBox log
\begin{equation}
	\vect{a}_{s} = \vect{a}_{e} \lor \vect{a}_{c},
\end{equation}
where $ \vect{a}_{e}$ and $\vect{a}_{c}$ are the feature vectors associated with the enterprise and CuckooBox samples, respectively, and $\vect{a}_{s}$ is the resulting synthetic feature vector that we use instead of the CuckooBox $\vect{a}_{c}$ vector in our validation. 

There is still a possibility that some of our detection results on the synthetic dataset are inflated because we could be using some CuckooBox features to increase our score (these features would not exist in deployment, so our true TPR would be potentially lower). So in addition to removing obvious CuckooBox features (see Feature Engineering), we also find all CuckooBox features that occur in $>1$\% of benign CuckooBox (in order not to select accidentally malware features due to mislabeled data), and we remove all of those features that have positive weights in our LR model. This gives us a very conservative guess for the TPR (though not a lower bound, since we potentially missed some features), and represents our best attempt at estimating the true deployment TPR vs. FPR tradeoff.

To sum it up, the random training testing split, the compile time split, and the family split, computed separately for CuckooBox and synthetic enterprise data form our six validations. In addition, our chances of detecting on environment is further minimized by our choice of a $\ell_1$-regularizer, since a feature that detects behavior would better generalize across our two environments, and so would likely be chosen over a feature that detects on just the environment.

\section{Results}
\label{sec:result}

Our experimental dataset consists of 32,078 samples and 6,898,953 unique extracted features.  Out of the samples, 17,399 are benign, 14,679 malicious. 20,362 audit logs are from binaries executed in CuckooBox, and 11,716 are Invincea's enterprise four minute windowed audit logs. Out of the 20,362 CuckooBox audit logs, 5,683 are benign and 14,679 are malicious. Of the 14,679 malicious audit logs, 3,010 are known to be malicious based on their origination. The density of the $A$ matrix, as computed as the number of non-zeros divided by number of entries, is $1.2\e{-4}$, indicating that the matrix is indeed very sparse. All computations were performed on a 2014 16GB MacBook Pro laptop running MATLAB 8.3. The LR classifier was computed using the Glmnet software package \cite{friedman2regularization,qian2013glmnet}, with 20-fold internal cross-validation for determining $\lambda$.

\begin{figure*}[ht]
	\center
	\includegraphics[width=1.02\textwidth]{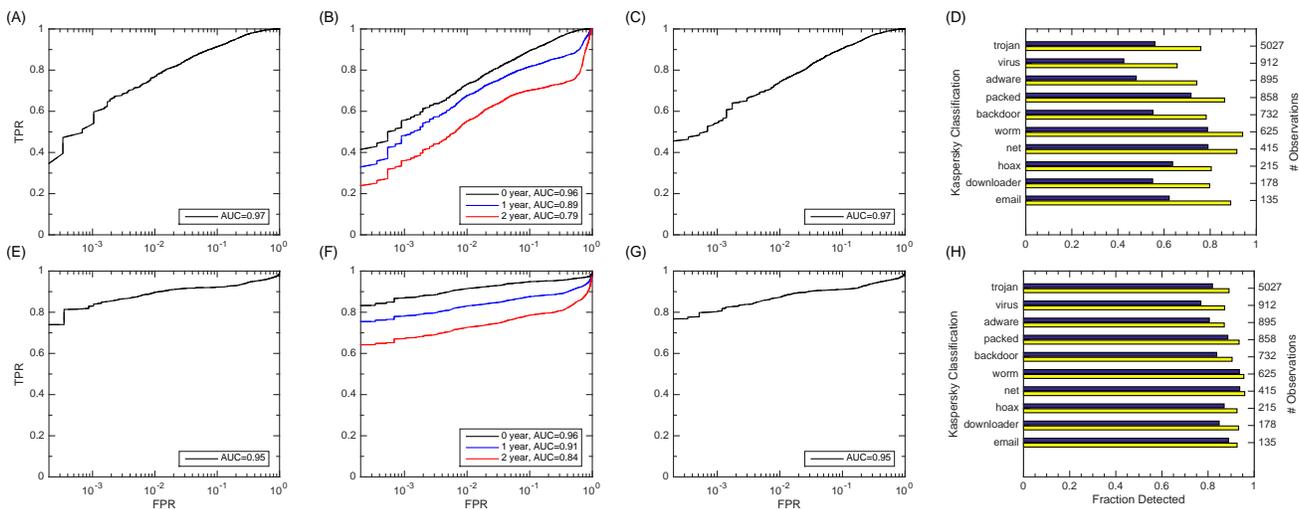}
   \caption{$10$-fold validation results for our $\ell_1$-regularized LR classifier. The top panels are validation results computed only on CuckooBox collected audit log data, while the bottom panels show validation results done only on synthetic malware and pure benign enterprise audit logs. Each column shares the same classifier between the top and bottom panel. The first column shows the ROC validation curves for randomly split data. The second column shows ROC validation curves based on compile time splits, where the training malware data is at least one (middle line) or two (bottom line) years older than validation malware data. The third column show ROC validation curved based on malware families splits, where we train on one set of malware families, and validate on the rest of the families. The fourth column shows results from the first column, at two deployment-relevant false positive rates ($10^{-2}$ top and blue, and $10^{-3}$ bottom and yellow), broken down by specific malware types, with the number of total observations of each malware type is shown on the right axis. }
   	\label{fig:validation}
\end{figure*}

The results for the six various validations of our LR classifier are shown in Fig. \ref{fig:validation}. In the top validation panels we test exclusively on CuckooBox data, while in the bottom panels exclusively on the synthetic enterprise data. Note that the classifier used to compute the TPR and the FPR in the top and bottom of each column is the same. Since we are proposing to deploy the classifier in an enterprise setting, we are primarily concerned with performances in the very low FPR regions ($\leq 10^{-2}$). We roughly estimate that the $10^{-2}$ and $10^{-3}$ FPR to be equivalent to a false alarm once a day and once every 8 days/ per computer, respectively, assuming 8 hr/day usage.

Fig. \ref{fig:validation}A shows that for random splitting our LR classifier has a robust detection level of $77$\% TPR at $1$\% FPR. Fig. \ref{fig:validation}E shows that for the synthetic enterprise version of the validation we can see the large improvement in the ROC curve at the lower FPR values due to the removal of the VirusTotal missed malware, with the TPR at about $89$\% for $1$\% FPR, and even larger gain at deployment-relevant FPRs. The associated bar plot of TPRs at FPRs $10^{-2}$ and $10^{-3}$ are shown in Fig. \ref{fig:validation}D and H, grouped by malware type (as determined by Kaspersky antivirus) .

In addition to the validation based on random dataset splitting, we also validated results by testing only on malware that is at least one or two years older than malware in the training set (Figs. \ref{fig:validation}B and F), and separately testing on malware families not included in our training dataset (Figs. \ref{fig:validation}C and G). The number of samples used to train the classifier in Fig. \ref{fig:validation}B and F is smaller than in the other panels, but all the ROC curves for all the years were computed using the same training dataset, with only the validation set being adjusted during ROC curve computation. The zero year ROC curve line is worse than the ROC curve in first column, even though it is the same validation, because the amount of training data is significantly reduced in the second column in order to be able to keep the same classifier for all three curves. 

We observe around an $8$\% drop in detection for each additional minimum year difference between the training and the testing set, which demonstrates that concept drift does affect our detection. The detection rate is still fairly high for the synthetic enterprise data (bottom panel), where the TPR is at least $67$\% with a FPR of $0.1$\%, and at least $73$\% for the FPR of $1$\%. These results are promising, considering that commercial antivirus solutions have a $60$\% TPR for previously unseen malware \cite{lastline2015}.

Since we did not filter all malware audit logs for executions that did not exhibit malicious behavior (ex. crashed prematurely or required interaction to activate), our reported TPR is potentially underestimated. We performed manual examination of a small fraction of the missed detections, and a large fraction of them where GUI installers that simply did not finish executing in an automated sandbox run. In terms of practicality, FPR is vastly more critical, since it controls the number of false alarms that a enterprise network administrator would have to handle. For example a detector with a $5$\% FPR is simply not deployable. Our results show that a significant level of detection can be achieved close to enterprise level FPRs.

\subsection{Detecting AV Missed Malware}

One important advantage of our approach over standard antivirus engines is that we are using observation vectors that are currently not utilized for detection, making it less likely that malware would be able to hide from it. As the result, assuming our classification labels are mostly correct, we are actually able to detect a large fraction of malware missed by popular antivirus engines. 

\begin{figure*}[ht]
	\center
	\includegraphics[width=0.80\textwidth]{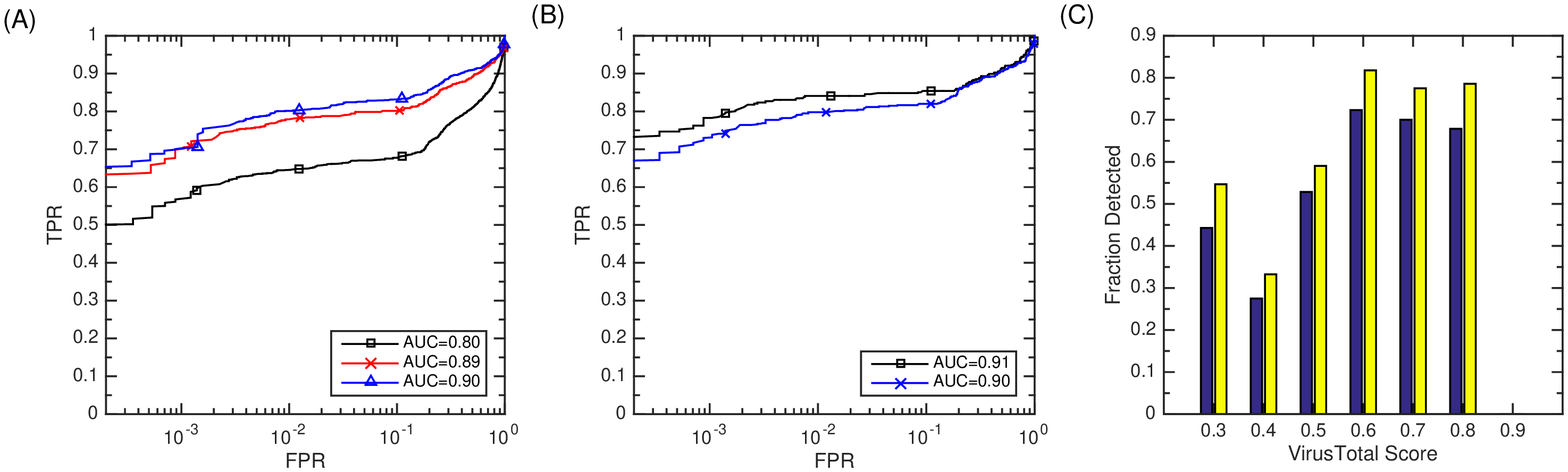}
   \caption{$10$-fold validated ROC curves of malware that is detected by our classifier and is missed by antivirus engines. In each instance the model was only trained on antivirus detected data, and validated only on malware missed by that engine. (A) Validation on specific antivirus engines. Kaspersky, black squares, McAfee red x, and Symantec, blue triangle. (B) Validation on meta engines. Kaspersky+McAfee+Symantec, black squares, McAfee+Symantec, blue x. (C) The fraction of malware detected, based on VirusTotal score for the composite (all three) engine, at two false positive rates ($10^{-2}$ left and blue, and $10^{-3}$ right and yellow). Lower VirusTotal scores indicate malware that is harder to detect}
   	\label{fig:av}
\end{figure*}

Recall that $88$\%, $95$\%, and $90$\% of our malware are detected by Kaspersky, McAfee, Symantec, as well as $96$\% when using McAfee and Symantec as a ``meta''-engine, and $98$\% when using all three. Removing those audit logs from our validation dataset, in Fig. \ref{fig:av} we show that we are able to detect a significant fraction of malware that is missed by standard antivirus engines, as well as meta-engines consisting of several popular antivirus engines. Specifically, we are able to detect around $80$\% of malware completely missed by a Mcafee + Symantec + Kaspersky ensemble detector.  Note that other than removal of detected malware from our validation dataset, the validation procedure in the figure is identical to that of Fig. \ref{fig:validation}E.

It is possible that some of our detection in Fig. \ref{fig:av} is due to the varying definition of malware between the major vendors and VirusTotal consensus. For example, while Kaspersky and Symantec have approximately the same detection rate on our dataset, our ROC curve for Kaspersky is worse than for Symantec. This suggests that Kaspersky has potentially a different threshold for what it considers malware, which results in our training data missing some category of malware.  Therefore, we have also analyzed our TPR for known malware, labeled as malware, not based on VirusTotal score, but due to known origination of the associated binary executable. Our analysis showed $79$\% TPR at $1$\% FPR, and $72$\% TPR at $0.1$\% FPR, when measuring performance on this known malware, using the Kaspersky malware only trained classifier, thus confirming our overall results.

\subsection{Important Features}

In addition to the validation above we also looked at the actual features that are used for detection. To do this, we recomputed the regularized LR model on our full dataset, which resulted in a LR model that uses 1,704 features. Of these features, the sum of the positive weights (indicating malicious features) was around 159 vs. -73 for the negative weights (indicating benign features). This shows that our detection relies more on malware behavior than benign behavior for classification, as we have expected.

While in general interpreting Windows audit logs is difficult without more fine grained knowledge (we only know the DLL executed, not the function called), we did observe some features that are directly interpretable. We note that none of the features by themselves are enough to classify a log as malicious, and that it is observation of at least several of these features that causes detection. To do this, we sort the non-zero LR features by their contribution to classification of malware, which we compute by multiplying the number of times a feature occurs in malware by the absolute weight in our LR model. Below are several top features that we found to have a direct interpretation:


The $\#2$ most important feature is:\\
\verb+Executing edited file+.\\
This represents an execution of a binary that was at some point before was edited by some process. While not always occurring in malicious software, this is clearly suspicious behavior.

The $\#3$ most important features is:\\
Write to \verb+...\windows defender\...\detections.log+.\\
This feature represents an actual detection by the Microsoft antivirus engine. This is a good validation of our algorithm, since this is clearly a strong non-signature indicator of malware, and we were able to discover it automatically.

The $\#9$ most important feature is:\\
Write to \verb+...[system]\msvbvm60.dll+.\\
While at first it might seem surprising that a library for Visual Basic 6 is an indicator of malware, a bit of research shows that the use of Visual Basic in malware has been on the rise because VB code is tricky to reverse engineer \cite{lavasoft2015}. Add to this fact that support for the language has ended more than 10 years ago, it is clear why this would be a good indicator of malware behavior.

While these are just three examples of the features that our LR model detects on, it clearly demonstrates that we are able to automatically recover some important non-signature events.

\subsection{Limitations}

It is clear that this, or any other approach, is not a panacea to the malware problem. Given a large enough deployment of such a detector, malware authors will start finding ways to obfuscate their audit log trail. For one thing, slow moving malware is an open problem, and potentially will not be detected using the four minute window of an audit log. 

The ROC curve estimates for the low FPR regions are not as reliable as for the less relevant higher FPR regions. More reliable estimates for the deployment relevant region would require significantly more samples. Also, given the somewhat artificial conditions in which we collected malware audit logs, our estimates of the TPRs could potentially be inaccurate. The TPR could further be affected by the constantly changing distribution of the malware in the wild, which we are not able to effectively approximate.

Our approach is only an initial demonstration of audit log detection feasibility and can undoubtedly be improved through a combination of better training data, improvement in feature engineering, and algorithmic tuning. Other problems can be mitigated by Microsoft improving their audit log capabilities. 

\section{Conclusion}
\label{sec:conclusion}

We demonstrated that audit logs can potentially provide an effective, low-cost signal for detecting malicious behavior on enterprise networks, and adds value to current antivirus detection systems. Our LR model yields a detection rate of $85$\% at an expected false positive rate of $1$\%, and detected $80$\% of malware missed by commercial antivirus systems. While audit logs do not directly record certain malicious tactics (\eg, thread injection), our results show that they still provide adequate information to robustly detect most malware. Since audit logs can easily be collected on enterprise networks and aggregated within SIEM databases, we believe our solution is deployable at reasonably low cost, though further work needs to be done to thoroughly test this claim.

Importantly, by putting multiple obstacles, such as audit log detection, in the way of would be attackers, we can increase the time and cost required to develop effective malware. In our future work we will explore improvements and integration of other audit log signals, like network flow, in our detection.

\section{Data and Software}

We are actively working on getting the full dataset anonymized, and approved for released to the public. The source code that completely reproduces our figures, and the link to our data can be found at GitHub:

\url{https://github.com/konstantinberlin/malware-windows-audit-log-detection}.
 
\section{Acknowledgement}

We would like to thank Robert Gove for his comments and discussion.

\sloppy

\bibliographystyle{abbrv}
\bibliography{pace}

\end{document}